\documentclass{aa}
\usepackage{lscape}
\usepackage{graphicx}
\usepackage{txfonts}
\def\bibl{\parindent=0pt \hangindent=0.150 in }

\begin{document}

\title{Search for Cold Debris Disks around  M-dwarfs}

\author{J.-F. Lestrade 
       \inst{1}
       \and
        M. C. Wyatt \inst{2}
       \and
       F. Bertoldi\inst{3}
       \and
       W. R. F. Dent\inst{4}
       \and
        K. M. Menten\inst{5}
}

\offprints{J.-F. Lestrade, e-mail : jean-francois.lestrade@obspm.fr}

\institute{Observatoire de Paris - CNRS, 77 av. Denfert Rochereau, F75014, Paris, France\\
\email{jean-francois.lestrade@obspm.fr}
\and
Institute of Astronomy, University of Cambridge, Cambridge, CB3 OHA, UK\\
\email{wyatt@ast.cam.ac.uk}
\and
Radioastronomisches Institut, Universit\"at Bonn, Auf dem H\"ugel 71, Bonn, D-53121, Germany\\
\email{bertoldi@astro.uni-bonn.de}
\and
UK Astronomy Technology Center, Royal Observatory, Observatory, Edinburgh, EH9 3HJ, UK\\
\email{dent@roe.ac.uk}
\and
Max-Planck-Institut f\"ur Radioastronomie, Auf dem H\"ugel 69, Bonn D-53121, Germany\\
\email{kmenten@mpifr-bonn.mpg.de}
}

\date{Received 22 june 2006 ; accepted 15 september 2006}

\abstract{Debris disks are believed to be related to 
planetesimals left over around stars after planet formation has ceased. The frequency of debris disks
around M-dwarfs which account for 70\% of the stars in the Galaxy 
is unknown while constrains have already been found for A- to K-type stars.
We have  searched for cold debris disks around 32 field M-dwarfs by conducting observations 
at $\lambda = 850\mu$m with the SCUBA   bolometer array camera at the JCMT and at   $\lambda = 1.2$mm 
with the MAMBO array at the IRAM 30-m telescopes. This is the first survey of a large sample of M-dwarfs
conducted to provide statistical constraints on debris disks around this type of stars.
We have detected   a new debris disk around 
the M0.5 dwarf GJ842.2 at $\lambda = 850\mu$m, providing evidence for cold dust at large distance 
from this star ($\sim 300$AU). By combining the  results of our survey with the ones 
of Liu et al. (2004), we estimate for the first time the detection rate of  cold debris disks 
around field M-dwarfs with ages between 20 and 200~Myr.  This detection rate  
is $13^{+6}_{-8}\%$  and is consistent with the detection rate  of  cold debris 
disks (9 -- 23\%) around  A- to K-type main sequence stars of the same age.
This is an indication that cold disks may be  equally  prevalent across stellar spectral types. 
\keywords{Stars : circumstellar matter; surveys; stars: low-mass; planetary systems : formation}}

\titlerunning{Search for Debris Disks around  M-dwarfs}
\authorrunning{Lestrade et al. }

\maketitle

%----------------------------------------------

\section{Introduction}
\label{Introduction}

In star-forming regions, a large fraction of the M-type low-mass stars younger 
than 6 Myr have disks of gas and dust that are primordial materials for
planet formation (Beckwith et al. 1990, Haisch et al. 2001, Dutrey et al. 1996,  
Andrews \& Williams 2005, Muzerolle et al., 2006). However, 
little is known about the frequency 
of planetary systems  around  mature M-type dwarfs (ages $> 10$ Myr)  
that account for  70 \% of the stars in the Galaxy. For example,
amongst the  $\sim$ 170 stars hosting planets 
discovered by the radial velocity technique, only three are M-dwarfs (GJ876 (Marcy et al. 2001),  
GJ436 (Butler  et al. 2004), and GJ581 (Bonfils et al. 2005).
This paucity may be due to selection effects because M-dwarfs are optically faint and chromospherically
active, making  radial velocity measurements more difficult than for solar-type stars (Endl et al. 2003).
From microlensing events, it has been infered that many M-dwarfs have planets with masses smaller than
Jupiter. However this inference is based on statistical arguments
from a small number of events (Gaudi et al., 2002 and Beaulieu et al., 2006).

Another approach to estimate the frequency of planetary systems around stars
is to measure the frequency of  debris disks around them.
A debris disk is made of planetesimals  left over from the planetary formation processes. 
In  these disks,   dust grains are continuously regenerated by collisions and/or  
evaporation of the planetesimals. This dust absorbs stellar radiation at visual wavelengths 
and reradiates the energy at infrared (IR) to submillimeter (sub-mm) wavelenghts.
It is the large emitting surface area of these numerous grains  that makes debris disks  
around stars observable in the IR and sub-mm, while the mass-dominant planetesimals remain undetected. 
IR to sub-mm excesses above photospheric emissions of  a few hundreds main sequence A- to K-type dwarfs
have been photometricly detected by IRAS (Aumann et al. 1984 ; Walker \& Wolstenscroft 1988 ; 
Backman \& Paresce 1993 ; Mannings \& Barlow 1998), by ISO (Habing et al. 2001 ; Spangler et al. 2001 ; Laureijs et al. 2002), 
and by Spitzer (Rieke  et al. 2005 ; Beichman et al. 2005 ; Chen et al. 2005 ; Bryden et al. 2006). 
 Futhermore, imaging of the emitting dust 
in the (sub)mm  of a few debris disks has revealed
substructures that are best interpreted as dust trapped in resonance with  an unseen planet in the inner part of the
system  (Wyatt 2003, 2006). Also the large reflecting surface area of the dust grains  
is  responsible for scattered light that provides a wealth of information 
useful to disclose the structure and dynamics of these systems 
({\it e.g.} $\beta$ Pic in Golimowski et al. 2006). Cold dust in the  Kuiper Belt and  warm dust in the 
asteroid belt are the debris of our solar system.

The first attempt to search for mid-IR excesses from field M-dwarfs, {\it i.e.} M-stars outside of star forming regions, 
yielded  only three detections in the IRAS point source catalogue (Inseok Song et al., 2002). 
However, debris disks around M-dwarfs may be missed by such mid-IR surveys because M-dwarfs 
are underluminous  ($L~=~0.1$ -- $0.001~L_{\odot}$)
so  that the irradiated dust is  cold ($<$20 K) and  more easily detected at (sub)mm wavelengths.  
In this paper, we report on observations of 32 young M-dwarfs at $\lambda= 850\mu$m with the 
Submillimeter Common User Bolometer Array (SCUBA) at the James Clerk Maxwell Telescope (JCMT),
and at $\lambda= 1.2$mm with the Max-Planck Millimeter Array (MAMBO) at the IRAM 30m telescope.
Previously, only three field M-dwarfs have been observed in the (sub)mm (Liu et al. 2004). 
Our aim is to determine  the frequency of debris disks around  M-dwarfs.

Several studies have adressed the problem of planet formation around M-dwarfs and their associated debris disks.
Mechanisms that might be specific to this stellar spectral type have been discussed by Johnstone  et al.  (1998),   
Laughlin  et al. (2004), Throop \& Bally (2005), Plavchan, Jura \& Lipscy (2005). Our contribution to establishing 
the  statistics of debris disks around M-dwarfs provides new observational constrains. 

Section 2 briefly describes the target stars.  Section 3  describes the observations at the  JCMT and IRAM30m (sub)mm telescopes. 
Section 4 describes the  results of these observations and provides details on the discovery of the debris disk
around the M0.5 dwarf GJ842.2.  We draw statistical conclusions
on debris disks around M-dwarfs in section 5.

\section {Target stars}

Observations have shown  that debris disks around main sequence stars of spectral types  A to K with ages less  
than 150 -- 400 Myr are dustier than older ones (Habing et al. 2001, Rieke et al.  2005). 
We therefore selected the youngest M-dwarfs in their post planetary formation phase with ages older than  10 Myr. 
We shall not denote them main-sequence stars  because an M0 star, {\it e.g.}, reaches the main sequence 
only after 900 Myr (Siess, Dufour \& Forestini 2000). 
We selected  the M-dwarfs that belong to the Moving Groups identified by Montes et al. (2001) 
and Zuckerman \& Inseok Song (2004a,b), since membership to such a group is a reliable age criterium.
The selected stars belong to the Local Association  open cluster (age = 20 -- 150 Myr),
IC2391 (35--55 Myr), the AB Dor moving group (100 -- 125 Myr as revised by
Luhman et al. 2005), the Castor moving group (200 Myr), the Ursa Maj moving group ($500 \pm 100$
Myr as revised by King et al.  2003), and the Hyades open cluster (600 Myr).  
Note that  HIP114066 is  part of the AB Dor group  according to Zuckerman \& Inseok Song (2004a,b)  
but is named GJ9809 and assigned to the  Local Association by Montes et al. (2001).In our surveys,
the completness for M-dwarfs in the Local Association is 61~\% (11 M-dwarfs observed/18 M-dwarfs in cluster), 
in Castor it is 75~\% (6/8), in Ursa Major it is 40~\% (4/10), in AB Dor it is 71~\% (5/7), 
in Hyades it is 29~\% (5/17), in IC2391 it is 50~\% (1/2). 
The M-dwarfs of these clusters that are not included in our surveys were inaccessible to the telescopes because 
of their low declinations. The selected stars have distances between 2.6 pc and  33.8 pc.
Our surveys at $\lambda=$1.2mm and 850$\mu$m  include
the observations of 32 different stars ; 12 were observed at the JCMT and 24 at the IRAM 30m telescope with four stars
in common in the two surveys (GJ285, GJ393, GJ9809, GJ4247).

\section {Observations}

\subsection  {JCMT/SCUBA observations}

The SCUBA bolometer array (Holland  et al. 1999) at the  JCMT in Hawaii at altitude 4092m
was used to observe  the 12 M-dwarfs listed in  Table~1. SCUBA uses two arrays of 37 and 91 bolometers to
simultaneously observe at 850$\mu$m  and 450$\mu$m, respectively,  the same region of the sky 
which is  $\sim 2.3'$ in size. The arrays are arranged as concentric hexagonal rings 
of bolometers at each wavelength. A fully sampled map of the $2.3'$ region 
can be made by ``jiggling'' the arrays according to an optimized pattern 
(Holland et al. 1999) but requires a long integration time to reach  the 2 mJy/$14''$beam sensitivity 
sought in our program. Standard photometry with a single bolometer of the array
({\it i.e.} on-off observations for sky background substraction)  yields sensitive observations 
in less time but the target source must be smaller than the beam (FWHM=$14''$ at 850$\mu$m). 
This is not certain for most of our targets because of the large {\it a priori} uncertainty on disk sizes (factor 3 - 5).  
Thus, in order to  optimize the use of our observing time, SCUBA data were taken in the 
 {\it wide photometry}  mode of  JCMT  described by Sheret, Dent and Wyatt (2004). 
It ensures that a source that is partially  resolved or offset from the observed position is detected 
with maximum sensitivity. In practice,
the arrays are moved through a 12-point pattern  that makes the central bolometer of each  array cycles 
through  4 inner points at  $1''$ from the target position and  8 outer points
at  $7''$. 

In general, an observation consisted of 78 integrations of 38 seconds, 
which corresponds to the total on+off time of 50 minutes. During each integration, 
the telescope beam cycled through the 12-point pattern described above and
the secondary mirror  continuously chopped the sky  over $60''$ at 7.8125~Hz in azimuth
to remove any large-scale sky variations. Each observation of a target was preceded
and followed by focusing and pointing. Atmospheric opacity $\tau_{\rm 225GHz}$ was measured  
by radiometry several times per minute and 
by ``skydips'' once every few hours ;   $\tau_{\rm 225 GHz}$, the zenith opacity at 225 GHz, the
operating frequency of the radiometer, was between 0.06 and 0.08 during our observations,
corresponding to precipitable water vapour columns of $\sim 1.2$ -- 1.6 mm.
The absolute flux density scale was checked by 11 measurements of standard calibrators at 850$\mu$m
(Uranus, Saturn, CRL2688, CRL618) over the  7  sessions of our program between August 26 2004 and October 07 2004.
From these measurements, we derived a conversion factor  FCF = $222 \pm 15$ (Jy/Volt)
for this period. The uncertainty in this factor provides  the accuracy of our absolute flux density scale, 
which is better than 10 \% at $850\mu$m. Observations at 450$\mu$m are $\sim$15 times less sensitive than at 850$\mu$m while
the flux density is only 6 times higher for  dust with a modified black-body
spectrum and $\beta=1$. None of our targets were detected at this shorter wavelength. 

The SURF package (Jenness et al. 2002) was used to combine the data from
different integrations, remove anomalous spikes, flatfield the array,
 subtract the sky-background noise level and apply atmospheric extinction corrections.
The calibrated data were then analysed by a specific software to average 
the data of the central bolometer and to produce
the map with the data of all the bolometers. This map is incomplete however because the sky is
undersampled in the wide photometry mode. Nonetheless it allows to realiably detect a source, 
although  interpretation of its measured flux density and morphology is somewhat complicated.
For the two sources detected (GJ842.2 and GJ696), we also re-analysed the data  with the Edinburgh custom data-reduction
software written in IDL and found similar  results.
The observations of the  12 M-dwarfs were conducted on 2004 August 26, 29, 31, 
on 2004 September 19 and on 2004 October 3, 6, 7. In the course of this programme, GJ842.2 
was observed 4 times, GJ82, GJ212, GJ696, and GJ890 twice and the other stars once.

 \subsection  {IRAM 30m/MAMBO observations}

   The 117-channel MPIfR bolometer array MAMBO-2 (Kreysa et al. 1998) at the IRAM 30-m telescope
on Pico Veleta at 2900m  altitude near Granada in Spain was used to observe the 24 M-dwarfs 
listed in Table~2. MAMBO-2 operates at an effective frequency 
of 250GHz ($\lambda$~=~1.20mm) with a half-power spectrum bandwidth of 80GHz. At $\lambda$~=~1.20~mm, 
the IRAM 30-m telescope has an effective beam of $10.7''$(FWHM), smaller than
the   angular sizes of the potential  debris disks  for most of our targets.
Simple on-off photometry  pointed  at the target  would therefore not be a
good strategy to detect a disk and multiple pointed observations, similar to 
wide photometry at JCMT,  would be inefficient at the IRAM 30-m telescope
because of the overhead due to  the motion of this large telescope. Thus, we mapped the targets.

We used the standard on-the-fly scanning mode, where the telescope scans in the azimuthal direction. 
The signal from the sky was modulated by the secondary mirror 
wobbling over a throw of $60''$ at 2 Hz in the scan direction (azimuth).
The wobble frequency  reflects a compromise between eliminating changes in the
atmosphere on as short a time-scale as possible and the challenges involved in moving  a 2-m secondary
at this frequency and keeping it mechanically stable. For each target, we made a map,
$400 \times 320$ arcsec$^2$ in size, scanned at a velocity of 4 arcsec s$^{-1}$ and with an elevation
spacing of 4 arcsec. This results in a fully-sampled map over this  area  with a $\sim$2mJy/$11''$beam 
rms noise in the central region per observation of 33 minutes (on+off times).  Some targets 
were observed twice or more in different scan directions and maps were co-added
to improve this rms noise. In practice, however, weather conditions for these multiple observations
were  different (opacity $ 0.2 < \tau_{\rm 225GHz} < 0.45$, low and medium sky-background noise)
and  thus, quite often, one of the maps dominates the final summation (except for GJ628).
The opacity of the atmosphere was measured every other hour by performing a skydip, but was also continuously
monitored with a radiometer located next to the telescope. Short on-off observations  of
position calibrators were carried out
before and after each observation of the programm sources to check the telescope pointing which was 
found to be excellent ($\sim 1''$). In order to set the absolute
flux density scale, primary flux calibrators (including planets) were observed at the beginning
and end of each run, which resulted in an absolute flux calibration uncertainty of $\sim 15 \%$ ($1 \sigma$).

\begin{table*}

\caption{JCMT/SCUBA wide photometry at 850$\mu$m.}

\begin{tabular}{l|lcclclccccc} 

\hline\hline\\

   Star    &   Sp.     &    RA$^a$   &   DEC$^a$  &  Dist.     & Log(L$_{x}$)$^b$ &   Moving      &  Integration &  Flux density          & Size $^c$ &   Dust    &   Dust      \\
           &   type    &    (J2000)  &   (J2000)  &   (pc)     & Log(ergs/s)     &   Group        &  time        &  (mJy)                 &  (AU)     & temp$^d$  &  mass$^e$   \\
           &           &             &            &            &                 &                & (hrs)        &                        &           &   (K)     &   (moon)   \\\\

\hline\\

 GJ82      &   M4      &  01 59 23.5 &  58 31 16  &   12.0     &   28.75         &   Loc Ass.     &   2          &    2.0  $\pm$ 1.4      &   84      &  16       &   $<$ 1.3  \\
 GJ212     &   M0.5    &  05 41 30.7 &  53 29 20  &   12.5     &  28.57          &  Loc Ass.      &   2          &    1.3  $\pm$ 1.4      &   88      &  23       &   $<$ 0.8  \\
 GJ285     &   M4.5    &  07 44 40.1 &  03 33 08  &   5.9      &  28.62          & Loc Ass.       &   1          &    -0.7 $\pm$ 1.9      &   41      &  20       &   $<$ 0.3  \\
 GJ393     &   M2      &  10 28 55.5 &  00 50 27  &   7.2      &  26.84          & Loc Ass.       &   1          &    1.9  $\pm$ 1.9      &   50      &  26       &   $<$ 0.3  \\
 GJ507.1   &   M1.5    &  13 19 40.1 &  33 20 47  &   17.4     &  n/a            & Loc Ass.       &   1          &    -0.4 $\pm$ 2.0      &  121      &  19       &   $<$ 2.8  \\
 GJ696     &   M0      &  17 50 34.0 & -06 03 01  &   21.9     &  n/a            & Loc Ass.       &   2          &     0.8 $\pm$0.8       &  150      &  20       &   $<$ 1.9  \\
 GJ9809    &   M0      &  23 06 04.8 &  63 55 34  &   24.9     &  29.43          & Loc Ass.       &   1          &   -5.2  $\pm$ 2.3      &  174      &  18       &   $<$ 7.6  \\
 GJ4247    &   M4      &  22 01 13.2 &  28 18 25  &    9.0     &  n/a            & Castor         &   1          &    1.1  $\pm$ 2.1      &   63      &  18       &   $<$ 0.9  \\
 GJ277B    &   M3.5    &  07 31 57.3 &  36 13 47  &   11.5     &  29.05          & Castor         &   1          &   -2.1  $\pm$ 1.8      &   80      &  17       &   $<$ 1.3  \\
 GJ842.2   &   M0.5    &  21 58 24.8 &  75 35 21  &   20.9     &  n/a            & Castor         &   4          &    {25  $\pm$ 4.6$^f$} &  300      &   13      &   $28 \pm 5$   \\
 GJ890     &   M2      &  23 08 19.5 & -15 24 35  &   21.9     &  29.22          &  Castor        &   2          &    -2.6 $\pm$ 1.6      &  153      &  16       &   $<$ 4.5  \\
 GJ875.1   &   M3      &  22 51 53.7 &  31 45 15  &   14.2     &  28.92          & IC2391         &   1          &    0.52 $\pm$ 2.1      &   99      &  17       &   $<$ 2.4  \\\\

\hline

\end{tabular}

\smallskip

\scriptsize

 $^a$  Observed positions, {\it i.e.} coordinates updated with the Hipparcos proper motions at the mean observation date of September 2004.   

 $^b$  ROSAT All Sky Survey of Nearby Stars : Huensch et al. 1999.   

 $^c$  Size is the radius of a debris disk as large as the telescope beam ($14''$).  

 $^d$  Temperature is computed at this radius.    

 $^e$  Mass upper limits are based on the $3\sigma$ flux densities (see text). 

 $^f$  Total flux density of the extended emission and the South-West peak.  

\end{table*}

The IRAM data were reduced using the MOPSIC software package (Zylka 1998). Data
were despiked, filtered for atmospheric variations, flat-fielded and corrected for
atmospheric opacity and turned into intensity maps and signal-to-noise ratio maps. 
The observations of the 24 M-dwarfs listed in Table~2 were conducted in  December
2004, in January, February and May 2005, and in January and March 2006.
Integration times were between 0.5 and 2 hours (see Table 2).
GJ628 has been a special case, after detecting possibly significant peaks in the first maps, 
we extended our observations to 20 hours of integration to reach
a map noise rms of 0.4mJy/$11''$beam in order to search for diffuse emission between these peaks.    

The relevant sensitivities for our SCUBA and MAMBO surveys are  in terms of circumstellar dust mass rather than 
mJy. Although the sensitivity  
is $\sim$~2~mJy/beam for each observations in both surveys,  the sensitivity  with MAMBO was
enhanced by taking a higher number of  observations and by averaging  the data over a canonical  disk area
assumed as large as the Kuiper-belt.  This averaging can be done also with the SCUBA 
wide photometry data but not as efficiently
because the sky  is not fully sampled. Also the dust mass sensitivity depends on
the stellar distance squared, and the M-dwarfs observed with MAMBO are closer than the ones observed 
by SCUBA in general. The all-in sensitivity is $\sim 1$ lunar mass of  circumstellar dust
for both surveys.

\section {  Results}

\subsection  {JCMT/SCUBA results}

\begin{figure}[t]
\centering
\includegraphics[width=9cm, angle=0]{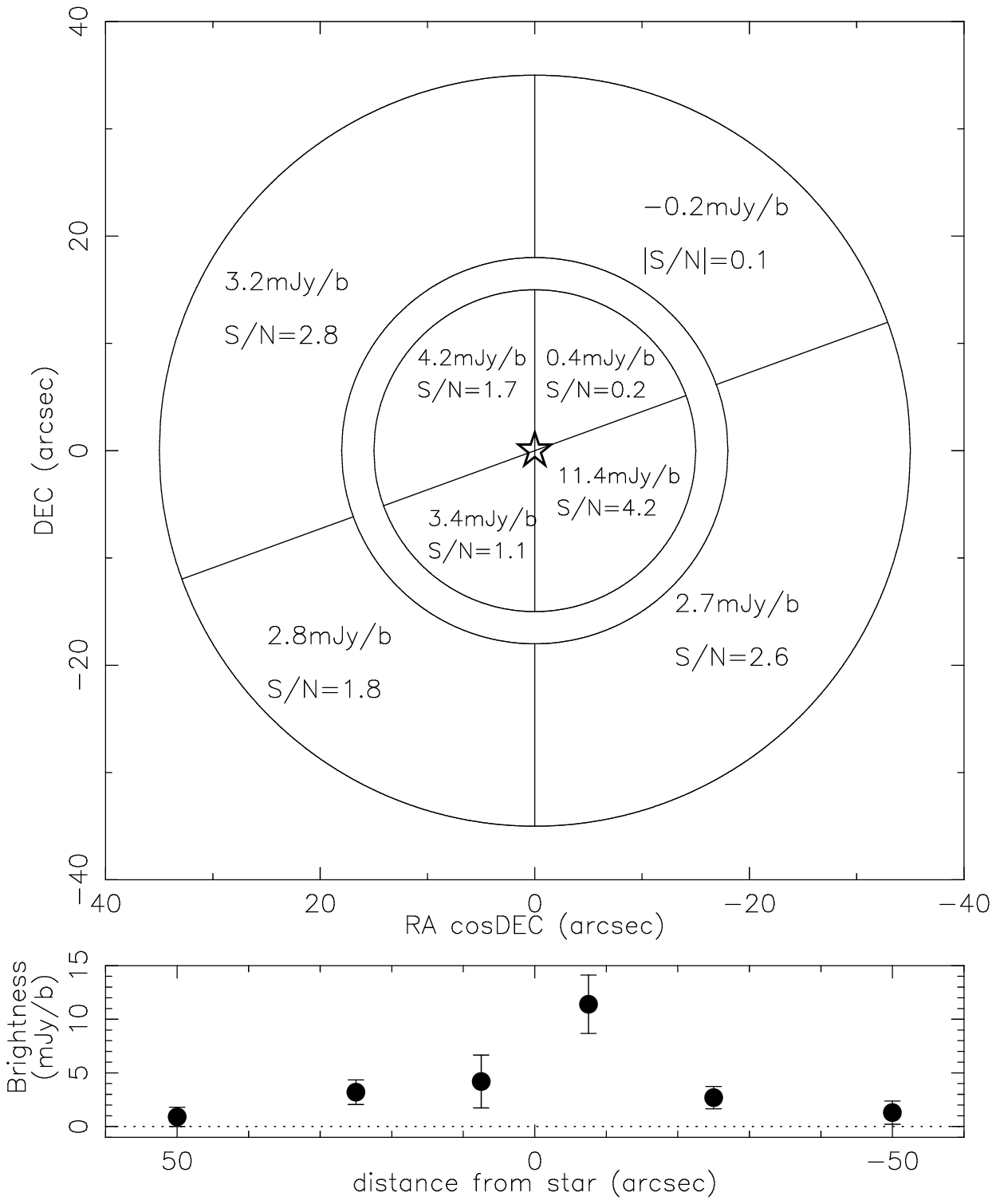}
\caption{
 Brightness distribution around the M0.5 dwarf GJ842.2 at $\lambda= 850\mu$m based on the data 
from all the bolometers of the JCMT/SCUBA array.  We made an attempt  to identify the general morphology of the source in splitting
the field of view in eight sectors and averaging the data in each one of them. As it is seen,
the emission is non-uniformly distributed and is oriented along the  North East - South West direction, suggesting an inclined disk. 
The plot at the bottom of the figure is a section of the  brightness along this direction.
The total flux density  (brightness $\times$ surface area) is estimated to be $\sim 25 \pm 4.6$mJy ({\it i.e.} S/N=5.5)
by adding the flux densities of the 4 quadrants along this direction.
The FWHM beamwidth of the telescope is $14''$ at 850$\mu$m. 
Observed position of GJ842.2 is $\alpha=$ 21h58m24.8s   and  $\delta= 75^{\circ} 35' 21''$  (J2000).  
}
\end{figure}

In our JCMT/SCUBA sample of 12 M-dwarfs, we have detected significant peaked and extended emissions 
around GJ842.2 revealing a debris disk around this M0.5 dwarf. We analysed the data of  GJ842.2 
in two ways. First, we straightforwardly averaged  the data of the central bolometer 
 and made a significant detection with a flux density of  4.8 $\pm$ 1.3 mJy.  
In effect, this flux density is the average of the emission 
at the 4 points $1''$ away from the star position and at the 8 points $7''$ away 
and may be underestimated  depending on the morphology of the source 
(Sheret, Dent and Wyatt, 2004). Nonetheless, the source is detected and this is the important point for
the survey. Second, we averaged the data from the first ring 
of bolometers (6 bolometers at mean radius $25''$ from center), and found
the significant mean brightness  of 2.4 $\pm$ 0.6 mJy/$14''$beam over this area.
We have also averaged the data of the second ring of bolometers further away  at $45''$ and  found $0.83 \pm 0.70$ mJy/beam, 
{\it i.e.}, consistent with no detection. The  third ring at $65''$ was used  
to remove sky-background noise and, thus, defines  the zero-level of the flux density scale.
This implies  that extended emission exists out to $\sim 25''$  and  so the  emitting  disk 
is as large as $\sim$ 500 AU at the distance of GJ842.2 (20.9~pc). 
In an attempt to identify the general morphology of this extended emission,  we 
split the field of view in eight sectors to average the data in each one of them (Fig.~1). 
In this decomposition, we used only two free parameters, the global orientation  of the sectors and 
the arclength of the inner South-West sector,  maximizing the signal-to-noise ratio 
in  it. The sector radii  are  not free parameters  but imposed by the SCUBA array. 
This averaging of the data indicates that the emission is non-uniformly
distributed and is oriented along the  North East - South West direction, suggesting an inclined disk. 
The plot at the bottom of Fig.~1 is a section of the  brightness along this direction.
The total flux density  (brightness $\times$ surface area) is estimated to be $\sim 25 \pm 4.6$mJy ({\it i.e.} S/N=5.5)
by adding the flux densities of the 4 sectors along the Norh East - South West direction. Finally,
we computed the likelihood that  the offset peak corresponding to the South-West sector  
($\sim 11$ mJy at $7''$) is a background source. The probability to find
a background source within a radius of $7''$ from an observed position in our 12 star survey is as low as 3\%,
computed from  $2100 \times (11/2.3)^{-1.5} \times \pi (7''/3600'')^2 \times 12$, based on the source count 
power-law $N(S) \propto S^{-1.5}$ normalized by 2100 sources/deg$^2$ for $\rm S > 2.3$~mJy at 850$\mu$m (Rowan-Robinson 2001).
So we conclude that there is a  cold debris disk around GJ842.2, possibly inclined, made of  a peak of emission 
at $7''$ South West from the star ($\sim$ 140 AU at 20.9pc) embedded in low brightness emission 
extending over $25 ''$ ($\sim 500$ AU). We emphasize that this  attempt to identify the general 
morphology of the disk is  speculative and that the straightforward averaging of the data 
of the first ring of bolometers mentioned above that indicates an extended source of 
mean brightness 2.4 $\pm$ 0.6 mJy/$14''$beam is our main result.

The total SCUBA flux density  $S_{\rm 850\mu{\rm m}}$ of GJ842.2 was used
to calculate the dust mass  $M_d$ of the disk by assuming a simple model 
of large grains ($100\mu$m in size) located at the mean
radius of the extended disk (300 AU). We use the standard optically 
thin dust formula for (sub)mm emission 
[$ S_{\lambda}=~{M_d~\times~B(\lambda,T_g)~\times~\kappa_{abs}}~/~{d^2}$, 
where  $T_g$ is the grain temperature, $d$ is the star distance, 
$\kappa_{abs}=~\kappa_{\rm 850\mu m}~{\big({850\mu m~\over~\lambda}\big)}^{\beta}$ 
is the mass opacity for the modified black body,
with $\kappa_{\rm 850\mu m}=~\rm 1.7~cm^2~g ^{-1}$ and $\beta=~0.8$ for  $\sim 100\mu$m  
size grains, Dent et al. (2000)]. At 300 AU, the dust temperature is only 13 K and 
the total dust mass is $28 \pm 5$ lunar masses for  $S_{850\mu m}=~25 \pm 4.6$ mJy.  
Those simple assumptions provide only the magnitude of the cold dust mass for the disk 
of GJ842.2  but it shows  clearly  that it is larger than the dust mass estimated
around the M1 dwarf  AU Mic ($\sim$ 1 lunar mass) and around the
M0.5 dwarf GL182 ($\sim 2.1$ lunar mass) by Liu et al. (2004).

In addition, we reanalysed  IRAS data of GJ842.2 with the on-line program {\it scanpi} and determined
the color-corrected flux density $57 \pm12$ mJy at 12$\mu$m. This flux density  
matches the photospheric emission computed from the  
NextGen model  for this M0 dwarf  ($\rm T_{photo} = 3500~K$, g = 5.5, [Fe/H] = 0.0)
by Allard et al. (2000, 2001),  normalized by  the B,V,R,I,J,H,K flux densities of  GJ842.2
accessible in the SIMBAD data base  (B, V bands : Third Catalogue of Nearby Stars (1991) by Gliese and Jahreiss, 
 Astron. Rechen-Institut, Heidelberg ; B, R, I bands : USNO-B1.0 catalogue, Monet et al. 2003 ; J, H,  K bands : 
2MASS All-Sky catalogue of point sources, Skrutskie et al. 2006). The scanpi 
signals in the other IRAS bands were not  significant
except possibly at 25$\mu$m  ($58 \pm13$ mJy, color corrected).
However, an IRAS flux density must be treated circumspectly 
when the signal-to-noise ratio is lower than 5.
If this 25~$\mu$m flux density ($4.5~\sigma$) were  real however, it would be clearly higher
than the photospheric level of GJ842.2 and so, interestingly,  indicative of warm dust close to the star. 
This needs confirmation  by new observations.

For the M0.5 dwarf GJ696 in our survey, the  central bolometer data did not yield
any significant flux density in the central region  close to this star (radius $< 14''$)
and data averaging of the  first ring  of bolometers $(18''-35'')$ did not reveal 
any extended emission  at the $1\sigma$ level of 0.8 mJy/$14''$beam. However,
we have  found a significant peak of emission, 11 mJy (4$\sigma$) at $20.8''$ 
and PA=+73$^{\circ}$ (East-North-East) from the  position of this star in the SCUBA incomplete map. It  
it is possible that its true flux density is underestimated because of  undersampling as mentioned above.   
This peak is 10 times higher than the noise of the first ring of bolometers.  
Thus, we do not regard this detection 
as evidence for a debris disk  around GJ696, referring to the SCUBA map of $\epsilon$~Eri 
which shows  ratio between  peaks and diffuse emission of less than 2 (Greaves et al. 2005). 
It is likely that the peak near GJ696 is a background source. We computed  
the likelihood  a sub-mm background source be within a radius of $21''$ over the whole 12 star sample 
using the source count at 850$\mu$m  by Rowan-Robinson (2001) and found a probability of 25~\%.
There is no corresponding 1.4GHz radio counterpart in the NVSS catalogue by Condon et al.  (1998) and no 
 optical object in the USNO-B1 catalogue that is positionally coincident. 

No peaked or diffuse emissions  have been detected around the other 10 stars of our survey. 
In Table 1, we derived the upper limits on the dust masses of their potential debris disks using 
the  $3\sigma$ flux density upper limits and the dust temperature for a linear separation from the star 
corresponding to  $7''$  which is the angular extent well probed by wide photometry.

\subsection{IRAM 30m/MAMBO results}

In our IRAM/MAMBO sample of 24 M-dwarfs (Table~2), we have detected 
emission in form of 5 isolated sources around the M3.5 dwarf GJ628 (Fig.~2). 
We have investigated whether or not these sources might simply be background 
galaxies. We have used the large catalogue USNO B1.0 (Monet et al. 2003) 
containing more that a billion objects with position uncertainties of $\sim 0.3''$ 
to search for optical counterparts and summarized our
results in Table~3. The only convincing association is  
for the source named MAMBO-W at $\sim 154''$ West of GJ628 in Fig.~2, based on coincidence between  MAMBO 
and USNO  positions at the $1\sigma$ level. In addition, this source MAMBO-W  matches within $\sim 1.7 \sigma$ the position
of the radio source NVSS163007-123940  in the 1.4~GHz NVSS catalogue by Condon et al.  (1998), if $\sigma$  is
the NVSS and MAMBO position uncertainties quadratically combined. Hence, we conclude that MAMBO-W is 
a background source identified in the visible and in the radio as  a quasi-flat spectrum source from cm (NVSS~:~$5.6\pm0.5$ mJy) 
to millimeter  wavelengths  (this paper~:~$\sim$ 3.6 mJy). For the other four MAMBO sources, 
the likelihood of optical associations is low (position discrepancies in Table~3)
and  they have  no  radio counterpart in the NVSS catalogue. 
Using the density of mm background sources, 600 sources deg$^{-2}$ 
for $\rm S_{\lambda} > 2$ mJy at 1.2~mm, found by Voss et al. (2006), we expect 1.1 background
sources  over the field that includes these four sources ($r \le 90''$, dotted circle in Fig~2). In this probability calculation,  
we did not  include the fields of the other 23 stars observed  because their  maps are not as deep as
the one of GJ628. Since we observe more MAMBO sources than statistically
expected from the sky background, they may be instead four dust clumps that are  part of a large disk (radius $\sim$ 400 AU)
surrounding the M3.5~dwarf GJ628. In this case, their  temperatures would be between 8 and 12K and their dust masses 
would be $\sim 0.1$ lunar mass each.  We found no significant mm diffuse emission  
between these point-like sources at the 0.4 mJy/$11''$beam level after
20 hours of integration. If  GJ628 were similar to  $\epsilon$~Eri, the expected diffuse component

\begin{landscape}

\begin{table}

%%\centering

\begin{center}

\caption{\scriptsize IRAM/MAMBO aperture photometry at $\lambda =1.2$mm.}

%%%{\scriptsize

\begin{tabular}{l|lcclclccccccc} 

\hline\hline\\

 Star      &   Sp.     &    RA  $^a$  &   DEC  $^a$  &  Dist.   & { Log(L$_{x}$)$^b$} & { Moving }  & {   Integr. }  & $\theta$  &  Mean brightness $^c$  & Map  noise  rms  & $F_{\theta}^d$     & Dust         &    Dust   \\
          &   Type     &  (J2000)     &  (J2000)     &  (pc)    & { Log(ergs/s)}          & {  Group }  &     time       &   ($''$)  & (mJy/11$''$beam)   & (mJy/11$''$beam) & $(3\sigma)$   & temp$^e$   &    mass $^f$  \\
          &            &              &              &          &                         &             &     (hrs)      &           &                    &                  &    (mJy)         &    (K)       &   (moon)  \\\\

\hline\\

GJ65A     &   M5.5     & {01 39 02.6}  & {-17 56 59} &   2.6    &     27.59     & {Hyades}    & 1.1    &  46    & -0.30  $\pm 0.5$  &  2.1       & 26  & 15     &   $<$ 0.9     \\
GJ109     &   M3       & {02 44 15.8}  & {25 31 22}  &   7.6    &     27.30     & {Hyades }   & 1.3    &  16    &  0.67  $\pm 1.4$  &  2.1       &  9  & 21     &   $<$ 1.6     \\
HIP16563B$^{g}$ &   M0       & {03 33 14.0}  & {46 15 19}  &  33.8    &     n/a       & {AB Dor}    & 0.9    &   4    &  0.35  $\pm 1.0^h$ &   -  &  3  & 27     &   $<$ 8.0  \\
HIP17695  &   M3       & {03 47 23.4}  & {-01 58 21} &  16.3    &     n/a       & {AB Dor}    & 0.5    &   5    & -0.79  $\pm 2.2$  &  2.2       &  7  &  20     &   $<$ 5.6      \\
GJ3379    &   M4       & {06 00 04.5}  & {02 42 23}  &   5.4    &     n/a       & {Hyades}    & 0.5    &  22    & -0.06  $\pm 1.1$  &  2.3       & 14  &  18     &   $<$ 1.5      \\
GJ234A    &   M4       & {06 29 23.6}  & {-02 48 53} &   4.1    &     n/a       & {Loc Ass}   & 1.1    &  29    &  0.76  $\pm 0.7$  &  1.8       & 14  &  18     &   $<$ 0.9       \\
GJ285     &   M4.5     & {07 44 40.0}  & {03 33 06}  &   5.9    &     28.62     & {Loc Ass }  & 1.1    &  20    & -0.13  $\pm 1.0$  &  1.7       &  9  &  17     &   $<$ 1.1       \\
GJ1111    &   M6.5     & {08 29 49.1}  & {26 46 29}  &   3.6    &     n/a       & {Castor }   & 0.5    &  33    & -0.34  $\pm 0.8$  &  2.4       & 22  &  12     &   $<$ 1.9       \\
GJ393     &   M2       & {10 28 55.3}  & {00 50 24}  &   7.2    &     26.84     & {Loc Ass }  & 1.7    &  18    & -0.38  $\pm 1.1$  &  1.5       &  7  &  24     &   $<$ 0.8       \\
GJ402     &   M4       & {10 50 51.8}  & {06 48 25}  &  6.8$^i$ &     n/a       & {Loc Ass}   & 1.1    &  18    & -0.24  $\pm 1.0$  &  1.7       &  8  &  18     &   $<$ 1.0      \\
GJ447     &   M4       & {11 47 44.5}  & {00 48 10}  &   3.3    &     26.84     & {Ursa Ma }  & 1.1    &  36    &  0.28  $\pm 0.4$  &  1.4       & 14  &  18     &   $<$ 0.6       \\
GJ408     &   M2.5     & {11 00 04.1}  & {22 49 57}  &    6.6   &    26.60      & {Castor }   & 1.1    &  18    &  0.16  $\pm 1.0$  &  1.7       &  8  &  22     &   $<$ 1.0       \\
GJ569A    &   M2       & {14 54 29.3}  & {16 06 03}  &   9.8    &    28.54      & {Ursa Ma }  & 1.1    &  12    &  1.76  $\pm 2.2$  &  2.2       &  7  &  24     &   $<$ 1.8     \\
GJ628     &   M3.5     & {16 30 18.0}  & {-12 39 51} &   4.3    &    26.48      & {Loc Ass}   & 20.0   &  28    &  0.1   $\pm 0.2$  &  0.4       &  3  &  12     &   $<$ 0.4      \\
GJ625     &   M1.5     & {16 25 24.8}  & {54 18 13}  &   6.6    &    26.70      & {Ursa Ma }  & 0.5    &  18    & -0.85  $\pm 1.5$  &  2.5       & 12  &  26     &   $<$ 1.3    \\
HIP81084  &   M0.5     & {16 33 41.5}  & {-09 33 12} &   31.9   &    n/a        & {AB Dor }   & 0.5    &   4    &  0.63  $\pm 2.6$  &  2.6       &  8  &  27     &   $<$ 19      \\
HIP86346  &   M0       & {17 38 39.6}  & {61 14 16}  &   24.5   &    n/a        & {AB Dor}    & 0.5    &   5    & -0.13  $\pm 2.1$  &  2.1       &  6  &  27     &   $<$ 8.2     \\
GJ791.2   &   M4.5     & {20 29 48.6}  & {09 41 19}  &   8.9    &    27.89      & {Hyades}    & 1.1    &  14    &  -0.98 $\pm 1.7$  &  1.7       &  6  &  17     &   $<$ 2.0     \\
GJ849     &   M3.5     & {22 09 40.7}  & {-04 38 26} &   8.8    &    n/a        & {Hyades}    & 0.5    &  14    &  -1.32 $\pm 2.7$  &  2.7       & 10  &  19     &   $<$ 2.7      \\
GJ876     &   M4       & {22 53 17.0}  & {-14 15 52} &   4.7    &    n/a        & {Loc Ass}   & 0.5    &  25    &  -0.07 $\pm 1.5$  &  2.8       & 19  &  18     &   $<$ 1.6     \\
GJ873     &   M3.5     & {22 46 49.3}  & {44 19 59}  &   5.0    &    29.09      & {Ursa Ma }  & 0.6    &  24    &  0.32  $\pm 1.1$  &  2.0       & 13  &  19     &   $<$ 1.1      \\
GJ4247    &   M4       & {22 01 13.2}  & {28 18 25}  &   9.0    &    n/a        & {Castor }   & 1.1    &  13    &  -0.18 $\pm 1.4$  &  1.4       &  5  &  18     &   $<$ 1.5    \\
GJ856A    &   M3       & {22 23 29.1}  & {32 27 32}  &   16.1   &    29.46      & {AB Dor }   & 1.1    &   8    &  -0.56 $\pm 1.5$  &  1.5       &  4  &  21     &   $<$ 3.3   \\
GJ9809    &   M0       & {23 06 04.8}  & {63 55 34}  &  24.9    &    29.43      & {Loc Ass }  & 0.5    &   7    &  0.36  $\pm 2.2$  &  2.2       &  7  &  26     &   $<$ 10  \\\\

\hline

\end{tabular}

%%}
\end{center}

%%%\smallskip

\scriptsize

 $^a$ Observed positions, {\it i.e.} coordinates updated with the Hipparcos proper motions at the mean observation date of July 2005.

 $^b$ ROSAT All Sky Survey of Nearby Stars : Huensch et al. 1999. 

 $^c$ Mean brightness  computed  over the effective area of angular size $\theta('')$ corresponding to the canonical disk 120 AU in diameter. 

 $^d$ Total flux density $F_{\theta}$ is the $3\sigma$ map noise integrated over this canonical disk.  

 $^e$ Dust temperature computed  at 60 AU from the central star.

 $^f$ Dust mass upper limits based on  $F_{\theta}$. 

 $^g$ Designation and spectral type from Zuckerman \& Inseok Song, 2004b and L\'opez-Santiago et al., 2006.

 $^h$ On-off single bolometer MAMBO standard photometry. 

 $^i$ Based on parallax from the Research Consortium on Nearby Stars RECONS project (www.chara.gsu.edu/RECONS).

\end{table}

\end{landscape}

\noindent should be seen at the $2 \sigma$ level in Fig.~2. Note that close inspection  
of this map does show that there is  a 1-$\sigma$ to 2-$\sigma$ emission extending between the E and NE MAMBO sources. 
It was not possible to image deeper with the MAMBO array in a reasonable oberving time to confirm such an  extended emission.  
Only additional observations,  in the Far-IR, in scattered light or by a long-term astrometric monitoring 
of the mm sources that would share the proper motion of GJ628 ($\sim 1.2''$/yr) if they are part of its debris disk, 
could provide decisive evidence for this intriguing result. 
We don't retain GJ628 as having a disk at this stage of our investigation.

We found the significant IRAS/{\it scanpi}  flux densities   $385 \pm 21$ mJy at 12$\mu$m 
and  $110 \pm 25$ mJy at 25$\mu$m (color-corrected) for GJ628, 
but they match the NextGen photospheric emission computed for this M3.5 dwarf. So 
no mid-IR emission excess above photospheric level is observed.

\begin{figure}
\centering
\includegraphics[width=7cm, angle=-90]{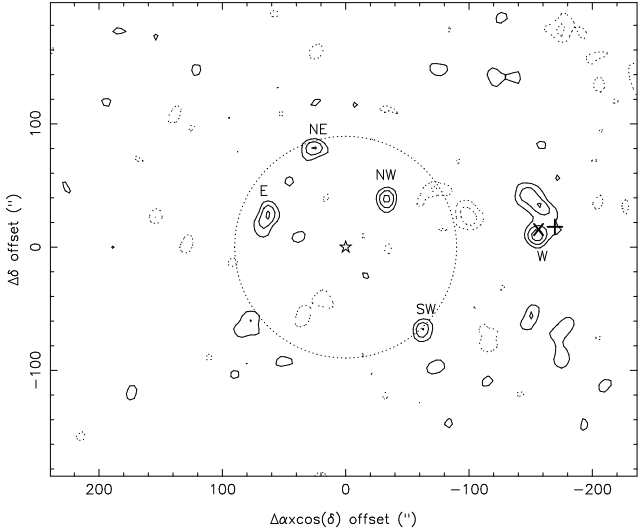}
\caption{IRAM/MAMBO Signal-to-Noise map of the field around the M3.5 dwarf GJ628 at $\lambda=1.2$mm. 
In this map, Signal-to-Noise ratio  (S/N) is computed from the intensity map 
reduced with the shift-and-add method using the software package mopsic (Zylka 1998) 
and smoothed to the angular resolution $12''$. The S/N is computed by dividing the intensity 
at each pixel by the local rms ($10 \times 10$ pixels) 
which is not uniform across this intensity map made by 
the scanning technique.  The  local rms $\sigma$ is $\sim$ 0.4 mJy/$11''$beam 
in the central region (r $ < 60''$) and degrades towards the edges of the map ; 
for example,  $1 \sigma$ is  $0.75$~mJy/$11''$beam at $ r \sim 150''$.
The countours are $-3 \sigma, -2 \sigma$ (dotted lines), 
and  $2 \sigma,   3 \sigma,  4 \sigma$;
The map pixel size is $3.5''$.  The  West component (MAMBO-W) is associated with 
a 1.4GHz radio source ($\times$ symbol) of  the NVSS catalogue and with an optical object 
of the USNO-B1.1 catalogue (+ symbol), and so it is a background source. 
The other four MAMBO components (NW, NE, E, SW) may be  dust clumps of a large debris disk 
around GJ628 but the lack of diffuse emission
between them requires additional observations for confirmation.
Observed position of GJ628 at center of the map is $\alpha=$ 16h30m18.0s and $\delta= -12^{\circ} 39' 51''$ (J2000).}
\end{figure}

For the other stars, no emission is apparent in their MAMBO maps, and no significant mean brightness is detected 
when averaging the data over  an effective disk with an  angular diameter  ($\theta$) corresponding to 
the canonical diameter 120 AU, {\it i.e.} the size of the Kuiper Belt. Hence,
we have used the map noise rms of each star
to derive the 3-$\sigma$ upper limit of its dust mass by averaging data over this canonical disk. 
In such a derivation, the $3 \sigma$  total flux density over the effective area ($\theta('')$) 
is $F_{\theta} = 3 \times {\rm rms} \times (\theta('')/11'')$ (mJy), which takes into account that 
the measured brightness uncertainty decreases 
as rms$/{(\theta / 11'')}$, {\it i.e.} as $\rm rms/\sqrt{\rm number~of~beams~in~the~disk~area}$,  while the total flux density
 increases as $(\theta / 11'')^2$, {\it i.e.} as $\rm {\rm (number~of~beams~in~the~disk~area)^2}$. 
The dust mass is  computed from this total flux density $F_{\theta}$ using  the optically thin dust model described above 
substituting  $F_{\theta}$ for $S_{\lambda}$ and taking the  absorption coefficient  
$ \kappa_{abs} = 1.3~{\rm cm}^2 {\rm g}^{-1}$ at $\lambda =1.2$~mm, 
according to the modified black-body law and $\beta=0.8$. 
The implicit assumption of a disk seen face-on in this derivation provides a conservative upper limit 
for the total flux density and hence for the circumstellar  dust mass since an inclined disk would 
be covered by a smaller number of beams.
In Table 2, we  report the values for  $\theta('')$, mean brightness, 
brightness uncertainty, map noise rms,  $F_{\theta}$ and dust mass upper limit. 
We do not confirm the $2\sigma$ flux density of 17~mJy  for GJ873 (EV~Lac)
at $\lambda= 1.1$~mm  reported by Doyle and  Mathioudakis (1991).

\begin{table*}

\begin{center}

\caption{Positions of the five IRAM/MAMBO sources  in the map of GJ628 at $\lambda = 1.2$mm (Fig.~2) and their 
nearest optical objects referenced in the USNO-B1.0 catalogue.}

\begin{tabular}{l|cccr|rr}

\hline\hline\\

     \multicolumn{1}{c}{}\vline      &                           \multicolumn{4}{c}{ MAMBO }\vline              &     \multicolumn{2} {c}{USNO-B1.0}     \\\\

\hline\\

     Name       &  $\alpha^a$        &   $\delta^a$                          &   Flux$^b$       & Star-source    & Object-source       &   Object~~~~~~   \\
                &  (J2000)           &  (J2000)                              &   density        & separation     & separation~~        &  magnitudes~~    \\
                &                    &                                       &   (mJy)          & ($''$) (AU)    &  ($''$)~~~~~~~~~~   &                  \\\\

\hline\\

     MAMBO-W    &  16 30 07.5      &  $-12$ 39 42   &   3.6      & 154~~~ 660  &  3.0 ~ (1.0$\sigma$) &  B=20.03 R=19.12       \\
     MAMBO-NW   &  16 30 15.6      &  $-12$ 39 11   &   2.7      & 53 ~~~230   & 13.3 ~ (4.0$\sigma$) &  B=20.50 R=17.90     \\
     MAMBO-SW   &  16 30 13.9      &  $-12$ 39 59   &   2.0      & 90 ~~~390   &  10.1~ (3.0$\sigma$) &  B=19.27 R=18.23       \\
     MAMBO-NE   &  16 30 19.7      &  $-12$ 39 32   &   2.4      & 83 ~~~360   &  9.1 ~ (2.5$\sigma$) &  B=20.06 R=18.30    \\
     MAMBO-E    &  16 30 22.3      &  $-12$ 39 25   &   1.8      & 68 ~~~290   &  6.9 ~ (2.0$\sigma$) &  B=21.63 ~I=18.61      \\\\

\hline

\end{tabular}

\smallskip

\scriptsize

 $^a$  The coordinates uncertainties are $3.5''$ for MAMBO.

 $^b$  Flux densities are the peak values unaffected by gaussian smoothing of the map.

\end{center}

\end{table*}

\section {Discussion  }

\subsection{Debris disks in our sample of 32 M-dwarfs }

Our surveys of 32 young M-dwarfs conducted at $\lambda = 850\mu$m 
and 1.2mm yielded the detection of one disk around the M0.5 dwarf GJ842.2.
The emission from this disk is a  peak  at $7''$ South West from
the star ($\sim$ 140 AU at 20.9pc) embedded in low brightness emission 
extending over $25 ''$ ($\sim 500$ AU). 
We fitted a simple dust model by assuming large grains ($\sim 100\mu$m) at mean disk radius 
(300 AU, 13K) that yielded  $28\pm 5$  lunar masses  of cold dust for the total flux density measured. 
No other map shows convincing evidence for a debris disk in our survey, 
although the SCUBA incomplete  map of the M0 dwarf GJ696 (21.9 pc) shows 
a localized  source ($11\pm 2.7$ mJy) at $21''$ from the star but no extended emission at 
the level of 0.8 mJy/$14''$beam over $35''$, and the MAMBO map of the M3.5 dwarf GJ628 (4.3 pc) shows
four mm sources within $90''$ (400 AU) from this star  but no significant diffuse emission at the 0.4 mJy/$11''$beam level.
We do not consider these isolated clumps as sufficient evidence  for 
debris disks around these last two stars. 
However, one of the most defining features
of debris disks that have been imaged in the (sub)mm is that they are not smooth, but clumpy, and so we
plan other observations to investigate these cases further.

It is interesting to discuss the lifetimes of grains in the debris disk of 
GJ842.2. Due to the low emission efficiency in the sub-mm of 
micron-sized grains, the $\lambda = 850\mu$m emission is expected to arise 
from grains larger than 100$\mu$m  (Wyatt \& Dent 2002).  Their 
 collisional lifetime  is $t_{coll} \ge$ 250 Myr and 
their Poynting-Robertson (P-R) drag lifetime  $\ge 22$ Gyr at distance $r \ge 140$ AU. 
These  lifetimes are larger than the star age (200 Myr) and so it means  
that these large grains  at $r \ge 140$ AU could be  original (``primordial'')  
from the early protoplanetary phase.  On the other hand,  grains closer to 
the star than 140 AU, if any,  could not be primordial because of their shorter $t_{coll}$, {\it i.e.}
more frequent  constructive or destructive collisions.  The large disk around 
GJ842.2 is thus unique  in terms of the possible presence of primordial 
and processed large dust grains at such a late stage of evolution and implies a large 
protoplanetary disk at birth.  
Previous to our survey, only the field M-dwarfs AU Mic and GJ182 were detected
in the sub-mm (Liu et al. 2004). These two stars are 10-20 Myr old (see Favata et al.  1998, 
and Barrado y Navascues et al. 1999  for the age of GJ182). 
The  disk of AU Mic was  resolved  by several workers (Kalas et al. (2004), Liu (2004) and Krist et al. (2005)) 
that found the  size $\sim$ 200 AU  smaller than the one for GJ842.2 while 
these two stars have  the same spectral type but one is twenty times  younger. 
Further investigations should attempt to establish whether the  environmental factor
or the evolutionary factor controls these two disks. 

Plavchan, Jura \& Lipsky (2005) have not detected any warm debris disks 
via $11 \mu$m excess in their sample of 9 M-dwarfs older than 10Myr.
These authors highlight that stellar wind drag  rather than  P-R drag 
is responsible for dispersal of the dust in these systems and this might explain the dearth
of debris disks around M-dwarfs. For GJ842.2 ($L_* = 0.063 L_{\odot}$),  a stellar wind with a mass-loss  rate $\dot M_{sw}$  20 times
the solar rate ($\dot M_{\odot} \sim 2~10^{-14}$ M$_{\odot}$/yr) would be sufficient to remove the primordial grains 
of its disk in 200 Myr according to  the P-R to stellar wind drag timescales ratio  ${t_{PR} / t_{sw}} \propto {\dot M_{sw} /  L_*}$ 
(equation~(3) of Plavchan, Jura \& Lipsky (2005)). In fact,  winds  with mass-loss rates 1000 times the solar rate 
are observed for M-dwarfs, especially when young,  and would make the existence of primordial 
dust after 200 Myr inconceivable. Unfortunately,  no X-ray or CaII data are available that 
would allow an estimate of $\dot M_{sw}$ for GJ842.2. However,  GJ842.2  is not a flare star, 
and its  $\rm H_{\alpha}$  absorption  (equivalent width $EW_{H_{\alpha}}=~-0.545~\AA$, 
in Gizis, Reid \& Hawley, 2002)  indicates a moderate  chromosphere and  magnetic activity, and thus  only a modest 
stellar wind (Mullan et al. 1992, 2001, and Wargelin \& Drake 2001).
Hence, one can speculate that GJ842.2 does not have a mass-loss rate larger than 20 times the solar rate and
its distant cold grains are unperturbed at age 200Myr.
Although GJ628 is not retained  as having a debris disk in the present 
analysis, it is  interesting to note that its  X-ray luminosity $(10^{26.79}~{\rm ergs}~{\rm s}^{-1}$ 
measured  by ROSAT, Schmitt \& Liefke (2004)) is low, similar
to that of the quiet Sun ($10^{26.5}~{\rm ergs}~{\rm s}^{-1}$), and its  $EW_{H_{\alpha}}$ ($-0.234~\AA$, 
in Gizis, Reid \& Hawley 2002) indicates a weak to moderate chromosphere. Hence, similarly to GJ842.2,  
 if GJ628 were surrounded by cold  grains, one can  speculate that they would be unperturbed at age 20-150Myr.
Finally, we note also that there is no  difference in disk detection rates between the subsample
of 11 stars with high L$_{x}$  ({\it i.e.} Log(L$_x$) $> 27.5$) 
and the subsample of 7 stars with low L$_{x}$ in Tables~1 and 2. 

We have critically reviewed the literature to list the properties of the debris disks around A- to K-type main
sequence stars in Table~4 and plotted their dust masses as a function of spectral types in Fig~3 after adding
our results for the M-dwarfs. The 200 Myr old debris disk around GJ842.2 is  relatively massive 
(28 lunar mass of dust) and  may be the tip of the iceberg for the M-dwarf disks. 
We note that this mass is comparable to the highest disk mass of 
the A- to K-type main sequence stars plotted in Fig.~3. Future surveys will have to establish  whether 
or not this  upper envelope for the non-detected M-dwarfs in Fig.~3 is a real envelope 
to assert whether or not disks around M-dwarfs are as massive as around A- to K-type stars.  Presently, 
this comparison is also complicated by the fact that the debris disks detected 
in the (sub)mm around these A- to K-type  stars  are from an IRAS-selected sample 
from essentially an all-sky survey, while the M-dwarfs in our analysis are from an unbiased (except for age) sample  
of 32 stars. Eventually, a broader comparison with
protoplanetary  disk masses and   central star masses (Natta, Grinin \& Mannings (2000) and  
Andrews \& Williams, 2005) will allow  to discuss  how the mass of the central star 
impacts planet formation.

\begin{table*}

\small

\begin{center}

\caption{Dust masses of debris disks around A- to K-type main sequence stars. Note that HD34700 and HD39944 
previously classified as debris disks have been rejected; HD34700 is a T Tauri (Sterzik et al. 2005)
and has gas emission (Dent et al. 2005) and HD39944 was confused with a background galaxy (Mo\'or et al. 2006).}

\begin{tabular}[h]{l|rccrcccc} 

\hline\hline\\

       Star           & Sp.  & dist   & age    &  $ \rm S_{\nu}$    & $\lambda$ & Disk   & Dust   &  Dust   \\
                      & Type &  (pc)  & (Myr)  & (mJy)              &  (mm)     & radius & temp   &  mass  \\
                      &      &        &        &                    &           &   (AU) & (K)    &  (moon)  \\     
\hline\\

      $\tau$ Ceti     & G8V  & 3.65   & 10000  &  $4.6 \pm 0.6^{a}$ &  0.85     & 55     & 40     &   0.04   \\
      $\epsilon$ Eri  & K2V  &  3.22  &730     & $40 \pm 1.5^{b}$   &  0.85     & 60     &  37    &   0.26  \\
      GL182           & M0.5 & 26.7   & 20-150 & $4.8\pm 1.2^{c}$   &  0.85     & 60     & 27     &   3.35  \\
%%    HD34700         & G0V  &  89    & n/a    & $40.7\pm 2.4^{d}$  &  0.85     &  60    &  49    & 148     \\
%%    HD39944         & G1V  &  109   & n/a    & $3.7\pm 0.9^{d}$   &  1.30     &  60    &  48    &  71     \\
      HD104860        &  F8  & 47.9   &  40    &  $6.8\pm 1.2^{d}$  &  0.85     &  60    &  49    &  7.1    \\
      HD8907          &  F8  & 34.2   &  180   &  $4.8\pm 1.2^{d}$  &  0.85     &  60    &  49    &  2.6    \\
      $\beta$ Pic     & A5V  & 19.3   & 12     & $104\pm 10^{e}$    &  0.85     &  60    &  77    &  11     \\
      HD38393         & F7V  & 8.97   & 1660   & $2.4\pm 1.0^{f}$   &  0.85     &  200   &  31    &   0.15  \\
      HD48682         & G0V  & 16.5   &  3300  & $5.5\pm 1.0^{f}$   &  0.85     &  110   &  38    &   0.9   \\
      HD74067B        & A2V  & 30     &  63    & $7.2\pm 1.5^{h}$   &  0.85     & 60     &  92    &   1.4   \\
      HD99803B        & A3V  & 100    & 123    & $4.7\pm 1.6^{h}$   & 0.85      & 60     &  86    &  11.3   \\
      HD107146        & G2V  & 28.5   &  80    & $20\pm 4^{i}$      & 0.85      & 60     &  47    &   7.8   \\
      $\eta$~Corvi    & F2V  & 18.2   & 1000   & $14.1\pm 1.8^{m}$  & 0.85      & 150    &  43    &   2.5   \\
      HD112412        & F1   & 25     & 28     & $3.8\pm 1.1^{h}$   & 0.85      & 60     &  64    &   0.8   \\
      HD128167        & F2V  &  15.5  &1700    & $6.2\pm 1.7^{f}$   &  0.85     & 320    &  31    &   1.2   \\
      HD131156        & G4V  &  6.70  & 200    & $6.2\pm 1.9^{j}$   &  1.30     & 60     &  45    &   0.4   \\
      Vega            & A0V  & 7.76   & 350    & $45.7\pm 5.4^{f}$  &  0.85     & 70     &  97    &   0.6   \\
      AU Mic          & M1   &  9.9   &12      & $14.4\pm 1.8^{c}$  &  0.85     & 40     &   31   &   1.2   \\
      Fomalhaut       & A3V  &  7.69  & 280    & $97\pm 5^{f}$      &  0.85     & 80     &  61    &   1.6   \\
      HD218396        & A5V  &  40    & n/a    & $28\pm 11^{k}$     & 1.10      & 60     &  77    &   24.6  \\\\

\hline

\end{tabular}

\smallskip

\scriptsize
 
 $^a$ Greaves et al. 2004 ;  $^b$ Greaves et al. 1998 ; 
 $^c$ Liu et al. 2004 ; $^d$ Najita \& Williams 2005 ; 
 $^e$ Zuckerman et al. 1993 ; $^f$ Sheret  et al. 2004 ; 

 $^h$ Wyatt et al. (2003) ;  $^i$  Williams et al. 2004 ;
  $^j$ Holmes et al. 2003 ; $^k$  Sylvester et al. 1996 ;  
 $^l$ Wilner et al. 2003 ; $^m$  Wyatt et al.  2005 

\end{center}
\end{table*}

\subsection {Detection rate of dusty debris disks around M-dwarfs}

We have combined our surveys (32 field M-dwarfs observed and 1 detection)  with  the JCMT survey of 3 field M-dwarfs
with 2 detections by Liu et al. (2004). 
Note that their small sample  has the same characteristics in ages, distances and sensitivity as  ours ; 
AU Mic (9.9 pc) and GJ799 (10.3 pc) of 
the  $\rm \beta~Pic$ moving group  (age 10 -- 20 Myr) and GJ182 (26 pc) of the Local Association.
In this combination, there are (1+2) detections out of (32+3) M-dwarfs searched,  yielding
the detection rate $f = 8_{-6}^{+4} \%$ for cold debris disks  in this sample of M-dwarfs with ages between 20 and 600~Myr
observed in the (sub)mm.
We have derived the uncertainty   by computing the  full width half maximum  of the 
 probability distribution  $P(k/N) = C_n^k(1-f)^{N-k}f^k$ that
$k$ detections be found in a sample of $N$ stars searched if the {\it a priori}
frequency of occurence is $f$. 

In order to compare  detection rates between surveys, it is important to compare observations
in the same wavelength regime since mid-IR surveys  probe warm dust ($\sim$~100K) 
close to the star ($\sim1$~AU)  while (sub)mm surveys  probe  cold dust ($\sim$20K)
distant from the star (100's~AU).
In the literature, there are two surveys of A- to K-type stars conducted in the  sub-mm 
that are suited for comparison with our detection rate. A sample of  A- and F-type main sequence stars
not detected by IRAS  with ages 10 -- 170 Myr was conducted  at $850~\mu$m 
by Wyatt, Dent \& Greaves (2003) and 
a sample of  F- to K- type stars with ages 10 -- 180 Myr 
was conducted at   $850~\mu$m by Najita and Williams (2005). 
It is important also to  trim the samples  so that they have same age interval since 
the decay of dust in debris disks with time seen for A-type  stars (Rieke et al. 2005) is likely to 
affect M-dwarfs too. For these considerations, we therefore
have restricted our (sub)mm sample to the (3+20) M-dwarfs with ages 20--200~Myr,  including the three M-dwarfs studied by 
Liu et al. (2004) satisfying  this age limit.  
We then find that our detection rate  $f$
becomes $13_{-8}^{+6} \%$ (3/23) for cold debris disks surrounding field M-dwarfs with ages between 20 and 200~Myr.
This detection rate $f$  compares well with 
the detection rate (9 -- 14\%) found in the   A- and F-type star sample  of  Wyatt, Dent \& Greaves (2003), and to
 the detection rate   
$23_{-15}^{+13}~\%$ (3 detections for 13 stars observed) in the  F- to K- type star sample 
 of Najita and Williams (2005). It is remarkable that these cold dust
detection rates  for M-dwarfs and  A- to K-type dwarfs 
match. This may indicate that the populations of cold debris disks among early and late-type stars are alike.
We have also considered  the sub-sample of 89 F- to K-type stars with 
ages between 10 and 300 Myr  observed at $\rm \lambda$ = 1.2mm, 2.7mm and 3mm by Carpenter et al. (2005) that
has a detection rate of $2\%$. It is somewhat lower than the rates above but
the longer wavelengths used in their observations would have required deeper observations to be directly comparable 
to the other surveys.

\begin{figure}[t]
\centering
\includegraphics[width=6.5cm, angle=-90]{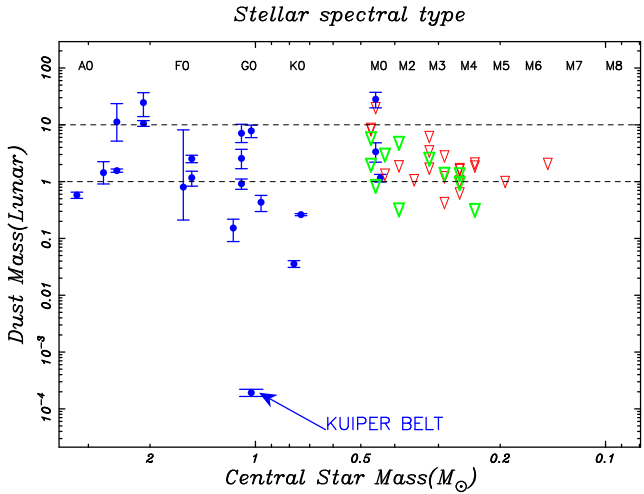}
\caption{
Dust masses of debris disks (filled circles) around A- to K-type main sequence stars (Table~4), around the
two M-dwarfs AU~Mic and GL182 from Liu et al. (2004) and around the M-dwarf GJ842.2  reported in this paper.
In addition,  we plot the dust mass $3\sigma$ upper limits of potential disks 
for the non-detections  of this paper and for the M-dwarf GJ799 reported in Liu et al. (2004) 
(red thin triangles= SCUBA and green thick triangles= MAMBO). The dust mass for
the Kuiper Belt is shown here to indicate that
present observations provide dust masses that are only the envelope of a possibly much larger population of debris disks.
The dashed lines are plotted as a guide.}
\label{FigVibStab}
\end{figure}

\section {Conclusions}

We have searched for cold debris disks around 32 field M-dwarfs 
at sub-mm and mm wavelengths. We report evidence for a new debris disk around the M0.5 dwarf  GJ842.2
of age  $\sim 200$~Myr which is massive ($28 \pm 5$ lunar masses) and large ($\sim 500$ AU). 
We report also the detection of four mJy  sources at 1.2mm around the M3.5 dwarf  GJ628  that
might be associated with a debris disk around this star unless they are background sources. 

By combining the result of our surveys with the two cold debris disks discovered by
Liu et al. (2004) in their observations of three field M-dwarfs, we derive for  the first time the
detection rate $f=13^{+6}_{-8}\%$ for cold debris disks
surrounding  M-dwarfs with ages 20--200 Myr. This detection rate  for M-dwarfs 
is  consistent with  the detection rate 9 -- 23\%  for cold disks around A- to K-type main-sequence  stars 
with ages 10 -- 180~Myr measured by  Wyatt, Dent \& Greaves (2003) and  Najita and Williams (2005)
in two  sub-mm surveys. 
The consistency between  these detection rates is 
the first indication that a   population of cold disks is equally  prevalent 
across stellar spectral types. The highest disk mass ($28 \pm 5$ lunar masses for GJ842.2) among the
 M-dwarfs is comparable to the highest disk mass among the  A- to K-type main sequence stars (Table~4 and Fig.~3). 
We plan additional observations of a larger sample of M-dwarfs to ascertain this detection rate and to better determine 
the dust mass upper envelope  of field M-dwarfs. The dust mass for the Kuiper Belt 
is $\sim 3~10^{-4}$ lunar mass (Landgraf et al., 2002, and  Moro-Mart\'in \&  Malhotra, 2003),
and, clearly, deeper observations are required to establish the statistics of debris disks 
around mature stars over  a large range of ages.

\begin{acknowledgements}

We are grateful to the staffs of the JCMT and of the IRAM 30-m telescope
for their  dedication in carrying out the
observations reported in this publication.  The JCMT project number was m04bu06.
This research has made use of the SIMBAD database and  of the VizieR catalog access tool,
operated at the Centre de Donn\'ees Stellaires (CDS), Strasbourg,
 France.  This publication makes use of data products from the Two
Micron All Sky Survey, from IRAS and Scanpi Processing maintained at
 the Infrared Processing and Analysis Center (IPAC),
California Institute of Technology, Pasadena, CA.

\end{acknowledgements}

\bibl
 Allard, F., Hauschildt, P. H., Alexander, D. R., Tamanai, A., \& Schweitzer, A.,  2001, ApJ, 556, 357
\filbreak

\bibl
 Allard, F., Hauschildt, P. H., \& Schwenke, D., 2000, ApJ, 540, 1005
\filbreak

\bibl
Andrews, S.M., Williams, J.P., 2005, ApJ, 631, 1134 
\filbreak

\bibl
Aumann, H. H., Beichman, C. A., Gillett, F. C.  et al, 1984, Ap.J., 278, L23-27
\filbreak

\bibl  
Backman D.E., Paresce, F., 1993, in Protostars and Planets III, ed. E.H. Levy and J.I. Lumine (Tucson : Univ. Press), 1253
\filbreak

\bibl 
Barrado y Navascués, D., Stauffer, J. R.,  Inseok Song,   Caillault, J.-P., 1999, ApJ, 520, L123
\filbreak

\bibl
Beaulieu, J.-P., et al., 2006, Nature, 439, 437 
\filbreak

\bibl  Beckwith, S.V.W., Sargent, A.I., Chini, R.S., G\"usten, R., 1990,  AJ, 99, 924
\filbreak

\bibl
Beichman, C.A., Bryden, G.,  Rieke, G.H.,  Stansberry, J.A., et al,  2005, ApJ, 622, 1160
\filbreak

\bibl
Beichman, C. A., Tanner, A., Bryden, G., Stapelfeldt, K. R., et al, 2006,  ApJ, 639, 1166
\filbreak

\bibl
Bonfils, X., Forveille, T., Delfosse, X., et al., 2005, A\&A, 443, L15
\filbreak

\bibl
Bryden, G., Beichman, C. A., Trilling, D. E.,  Rieke, G. H., et al, 2006, ApJ, 636, 1098 
\filbreak

\bibl
 Butler, R.P., Vogt, S.S., Marcy, G.W., Fischer, et al., 2004, ApJ, 617, 580
\filbreak

\bibl
 Carpenter, J.M. et al.,  2005, AJ, 129, 1049
\filbreak

\bibl
Chen, C. H., Patten, B. M.,  Werner, M. W., Dowell, C. D., et al., 2005, ApJ, 634, 1372
\filbreak

\bibl
Condon, J.J, Cotton, W.D., Greisen, E.W., et al., 1998, AJ, 115, 1693-1716
\filbreak

\bibl
Dent, W. R. F., Walker,  H. J. ,  Holland, W.S.,  \& Greaves, J.S., 2000,
M.N.R.A.S., 314, 702
\filbreak

\bibl
Dent, W. R. F., Greaves, J.S., Coulson, I.M., 2005, M.N.R.A.S., 359, 663
\filbreak

\bibl
Doyle, J.G. \& Mathioudakis, M., 1991, A \& A, 241, L41
\filbreak

\bibl
Dutrey, A., Guilloteau, S., Duvert, G., Prato, L;, Simon, M., Schuster, K., M\'enard, F., 1996, A\&A, 309, 493  
\filbreak

\bibl
Endl, M., Cochran, W.D., Tull, R.G., MacQueen, P.J., 2003, AJ, 126, 3099
\filbreak

\bibl
Favata, F., Micela, G., Sciortino, S., D'Antona, F, 1998, A\&A, 335, 218
\filbreak

\bibl
Gaudi, B.S., et al., 2002, ApJ, 566, 453 
\filbreak

\bibl
Gizis, J.E.,  Reid, I. N., \& Hawley, S.L., 2002, AJ, 123, 3356
\filbreak

\bibl
Golimowski, D. A., Ardila, D. R., Krist, J. E., Clampin, M., Ford, H. C. et al., 2006, AJ, 131, 3109
\filbreak

\bibl
Greaves, J.S. et al., 1998, ApJ, 506, L133
\filbreak

\bibl
Greaves, J.S.  et al., 2004, MNRAS, 351, 54
\filbreak

\bibl
Greaves, J. S., Holland, W. S., Wyatt, M. C., Dent, W. R. F., et al, 2005, ApJ, 619, 187
\filbreak

\bibl
Habing, H.J.,  Dominik, C., Jourdain de Muizon, M.,   Laureijs, R.J., et al., 2001, A\&A, 365, 545
\filbreak

\bibl
Haisch, K.E., Lada, E.A., Lada, C.J., 2001, ApJ, 553, L153
\filbreak

\bibl
Holland, W.S. et al, 1998, Nat., 392, 788-791
\filbreak

\bibl
Holland, W.S., et al, 1999, MNRAS, 303, 659
\filbreak

\bibl
Holland, W.S., et al, 1999, MNRAS, 303, 659
\filbreak

\bibl
Holmes, E. K., Butner, H.M., Fajardo-Acosta, S.B., Rebull, C.R.,  2003, AJ, 125, 3334
\filbreak

\bibl
Huensch M., Schmitt J.H.M.M., Sterzik M.F., Voges W.,  1999, Astron. Astrophys. Suppl. Ser. 135, 319
\filbreak

\bibl
Inseok Song,  Weinberger, A. J., Becklin, E. E.,  Zuckerman, B., Chen, C.,  2002, AJ, 124, 514-518
\filbreak

\bibl
  Jenness, T., Stevens, J.A., Archibald, E.N., Economou, F.,
  Jessop, N.E., Robson, E.I., 2002, MNRAS, 336, 14
\filbreak

\bibl
Johnstone, D., Hollenbach, D., Bally, J., 1998, ApJ, 499, 758
\filbreak

\bibl
Kalas, P., Liu, M.C., Matthews, B.C., 2004, Science, 302, 1990
\filbreak

\bibl
King, J.R., Villarreal, A.R., Soderblom, D.R., Gulliver, A.F., Adelman, S.J., 2003, AJ, 125, 1980-2017
\filbreak

\bibl
Kreysa, E. et al., 1998, Proc. SPIE, 3357, 319
\filbreak

\bibl 
Krist, J.E., et al., 2005,  A.J., 129, 1008
\filbreak

\bibl
Landgraf, M. et al., 2002, AJ, 123, 2857
\filbreak

\bibl
Laughlin, G., Bodenheimer, P \& Adams, F.C.,  2004, ApJ, 612, L73
\filbreak

\bibl
Laureijs, R.J., Jourdain de Muizon, M., Leech, K., et al., 2002, A\&A, 387, 285 
\filbreak

\bibl
Liu, M. C. et al., 2004, ApJ, 608, 526
\filbreak

\bibl
Liu, M.C. 2004, Sci, 305, 1442
\filbreak

\bibl
L\'opez-Santiago, J., Montes, D., Crespo-Chacon, I., Fern\'andez-Figueroa, M.J., 2006, ApJ, 643, 1160.
\filbreak

\bibl
Luhman, K.L., Stauffer, J.R., Mamajek, E.E., 2005, ApJ, 628, L69-L72
\filbreak

\bibl
Mannings, V. \& Barlow, M.J., 1998, ApJ, 497, 330
\filbreak

\bibl
 Marcy, G., Fischer, D., Vogt, S.S., Lissauer, J.J., Rivera, E.J.,  2001, ApJ, 556, 296
\filbreak

\bibl
Monet, D.G., Levine S.E., Cazian B., et al., 2003, Astron. J., 125, 984 
\filbreak

\bibl
Montes, D., L\'opez-Santiago, J., G\'alvez, M. C., et al., 2001, MNRAS, 328, 45
\filbreak

\bibl
Mo\'or, A., Abrah\'am, P., et al.,  2006, ApJ, 644, 525
\filbreak

\bibl
Moro-Mart\'in, A., Malhotra, R., 2003, AJ, 125, 2255
\filbreak

\bibl
 Mullan, D.J., MacDonald, J., 2001, ApJ, 559, 353
\filbreak

\bibl
Mullan, D.J. et al, 1992, ApJ, 397, 225
\filbreak

\bibl
Muzerolle, J., Adame, L., D'Alessio, P., Calvet, N., Luhman, K. L., et al., 2006, ApJ, 643, 1003 
\filbreak

\bibl
Najita, J., Williams, J.P., 2005, ApJ, 635, 625
\filbreak

\bibl
Natta, A., Grinin, V., Mannings, V.,  2000, in Ptotostars and Planets IV, Book - Tucson : University of Arizona Press, 559
\filbreak

\bibl
Plavchan, P., Jura, M.,  \& Lipscy, S.J., 2005, ApJ,631,1161
\filbreak

\bibl
Rieke, G. H., Su, K. Y. L., Stansberry, J. A., Trilling, D., et al., 2005, ApJ, 620, 1010.
\filbreak

\bibl
Rowan-Robinson, M., 2001, ApJ, 549, 745
\filbreak

\bibl
Schmitt, J. H. M. M. \&  Liefke, C., 2004, A\&A, 417, 651
\filbreak

\bibl
Sheret, I., Dent, W.R.F., Wyatt, M.C., 2004, MNRAS, 348, 1282-1294
\filbreak

\bibl
Siess, L.,  Dufour, E.  \& Forestini, M., 2000, A\&A, 358,593
\filbreak

\bibl
Skrutskie, M.F.,  Cutri, R.M., Stiening, R., Weinberg, M.D. et al,  2006, AJ, 131, 1163.
\filbreak
\bibl

\bibl
Spangler, C., Sargent, A.I., Silverstone, M.D., Becklin, E.E., \& Zuckerman, B., 2001, ApJ, 555, 932.
\filbreak

\bibl
Sterzik, M.F., Melo, C.H.F., Tokorinin, A.A., van der Bliek, N., 2005, A \& A, 434, 67
\filbreak

\bibl
Sylvester, R.J., Skinner, C.J., Barlow, M.J., Mannings, V., 1996, MNRAS, 279, 915.
\filbreak

\bibl
Sylvester, R.J., et al, 2001, MNRAS, 327, 133.
\filbreak

\bibl
Throop, H.B., Bally, J., 2005, ApJ, 623, L149.
\filbreak

\bibl
Voss, H., Bertoldi, F., Carilli, C., Owen, F. N., et al., 2006,   A\&A, 448, 823
\filbreak

\bibl
Wargelin, B.J., Drake, J.J., 2001, ApJ, 546, 57
\filbreak

\bibl
Williams, J.P. et al, 2004, ApJ, 604, 414
\filbreak

\bibl
Wilner, D.J., Holman, M.J., Kuchner, M.J., Ho, P.T.P., 2003, ApJ, 596, 597
\filbreak

\bibl
Wyatt, M.C., Dent, W.R.F., 2002,  MNRAS, 334, 589
\filbreak

\bibl
Wyatt, M.C., Dent, W.R.F., Greaves, J.S., 2003,  MNRAS, 342, 876
\filbreak

\bibl
Wyatt, M.C., 2003,   ApJ, 598,  1321
\filbreak

\bibl
Wyatt, M.C., Greaves, J.S., Dent, W.R.F., Coulson, I.M., 2005, ApJ, 620, 492
\filbreak

\bibl
Walker, H.J., Wolstencroft, R.D., 1988, PASP, 100,  1509. 
\filbreak

\bibl
Zuckerman, B.,  Becklin, E.E., 1993, ApJ, 414, 793
\filbreak

\bibl
Zuckerman, B., Inseok Song,  2004a, ARA\&A, 42, 685
\filbreak

\bibl
Zuckerman, B., Inseok Song, 2004b, ApJ, 613, L65
\filbreak

\bibl
Zylka, R.,  http://www.iram.es/IRAMES/otherDocuments/ manuals/Datared/
\filbreak

\end{document}